\documentclass[twocolumn,prb,aps,superscriptaddress,showpacs]{revtex4}
\usepackage{graphicx}

\begin{document}


\title{Spin-resolved impurity resonance states in electron-doped
  cuprate superconductors}

\author{Bin Liu$^*$}
\affiliation{Max-Planck-Institut f\"ur Physik komplexer Systeme,
D-01187 Dresden, Germany}

\begin{abstract}

With the aim at understanding the non-monotonic
$d_{x^{2}-y^{2}}$-wave gap, we analyze the local electronic
structure near impurities in the electron-doped cuprate
superconductors. We find that the local density of states near a
non-magnetic impurity in the scenario of $d_{x^{2}-y^{2}}$-wave
superconductivity with higher harmonics is qualitatively different
from that obtained from the $d_{x^{2}-y^{2}}$-wave superconductivity
coexisting with antiferromagnetic spin density wave order. We
propose that spin-polarized scanning tunneling microscopy
measurements can distinguish the two scenarios and shed light on the
real physical origin of a non-monotonic $d_{x^{2}-y^{2}}$-wave gap.

\end{abstract}

\maketitle


Pairing symmetry in the electron-doped cuprate high-temperature
superconductors has been extensively studied experimentally and
theoretically. In contrast to the hole-doped cuprates, where the
$d_{x^{2}-y^{2}}$-wave pairing symmetry has been generally
accepted\cite{krotkov,manske}, the earlier point contact tunneling
spectra suggested an s-wave like symmetry due to the absence of zero
bias conductance peak in the spectrum\cite{alff,biswas}. Recently,
the phase sensitive scanning (SQUID) measurements\cite{tsuei},
nuclear magnetic resonance study\cite{zheng}, and angle-resolved
photoemission spectroscopy (ARPES) experiments\cite{matsui,arm} have
provided strong evidences that the electron-doped cuprates are also
the $d_{x^{2}-y^{2}}$-wave superconductors. However, the functional
form of the $d_{x^{2}-y^{2}}$-wave gap in electron-doped materials
is a more subtle issue. A non-monotonic $d_{x^{2}-y^{2}}$-wave gap
with a maximal value in between nodal and antinodal points on the
Fermi surface (FS) has been measured in Raman experiments in
NCCO\cite{blumberg} and ARPES data on the leading-edge gap in
Pr$_{0.89}$LaCe$_{0.11}$CuO$_{4}$\cite{matsui}.

Up to now, the physical origin of such non-monotonic
$d_{x^{2}-y^{2}}$-wave gap is still under debate. Two kinds of
theoretical explanations have been put forward. One is to extend the
superconducting (SC) gap out of the simplest $d_{x^{2}-y^{2}}$-wave
via the inclusion of higher harmonics\cite{krotkov,manske}. Based on
the theoretical assumption, the $d_{x^{2}-y^{2}}$-wave pairing is
caused by the attractive interaction with the continuum of
overdamped antiferromagnetic (AF) spin fluctuations, which generates
a maximal gap near the hot spots (the points along the FS separated
by the AF move vector $Q_{AF}$). Since the hot spots in the
optimally doped NCCO and PCCO are located close to Brillouin zone
diagonals, one can generally expect the $d_{x^{2}-y^{2}}$-wave gap
to be non-monotonic. The other one is the coexisting scenario in
which the AF long-range order coexists with the
$d_{x^{2}-y^{2}}$-wave order\cite{yoshi,yuan}. The neutron
scattering\cite{kang} and transport experiments\cite{lavrov} have
observed a robust AF order, which survives a broad doping region in
the phase diagram. On the other hand, the ARPES measurements
revealed the intriguing doping evolution of the FS in
NCCO\cite{armitage}, where two inequivalent pockets around $(\pi,0)$
and $(\pi/2,\pi/2)$ shown in the FS have been explained to the band
folding due to the AF order\cite{luo}. As a consequence, the
resulting quasiparticle excitation can be gapped by both orders, and
the non-monotonic $d_{x^{2}-y^{2}}$-wave gap appears naturally.

In this paper, we argue that the local electronic structure near
impurities can provide important insight into the physical origin of
non-monotonic $d_{x^{2}-y^{2}}$-wave gap. We calculate local density
of states (LDOS) around a non-magnetic impurity starting from two
scenarios: $d_{x^{2}-y^{2}}$-wave superconductivity with a higher
harmonic versus $d_{x^{2}-y^{2}}$-wave superconductivity coexisting
with AF spin density wave (SDW) order. We find that the behavior of
density of states (DOS) in both scenarios suggests the presence of a
non-monotonic $d_{x^{2}-y^{2}}$-wave gap. Taking the single
non-magnetic impurity into account, we find that in the scenario of
$d_{x^{2}-y^{2}}$-wave superconductivity with a higher harmonic, the
LDOS behaves similar to that shown in hole-doped
cuprates\cite{morr}, $i.e.$ a $single$ resonance state near Fermi
energy appears at impurity site. However, due to introducing AF SDW
order in the latter scenario, the LDOS indicates a spin-resolved
feature, $i.e.$ two resonance states occur at impurity site with
different energies. For the sufficiently large SDW order, one spin
component (spin-up or spin-down) turns out to be dominant, and
although the DOS shows a U-shaped behavior, the presence of
resonance states at low energies in LDOS rules out the possibility
of s-wave pairing symmetry\cite{alff,biswas,yu,liu1}. Thus, we
conclude that the different electronic structure near a non-magnetic
impurity can differentiate between above scenarios and can be
checked by the further scanning tunneling microscopy (STM)
experiments.


We start from a phenomenological superconducting Hamiltonian
$H_{SC}=\sum_{{\bf k}\sigma}[\xi_{\bf k}c_{{\bf
k}\sigma}^{\dagger}c_{{\bf k}\sigma}+\Delta_{\bf k}(c_{{\bf
k}\uparrow}^{\dagger}c_{-{\bf k}\downarrow}^{\dagger}+c_{-{\bf
k}\downarrow}c_{{\bf k}\uparrow})]$, where $c_{{\bf
k}\sigma}^{\dagger}$ ($c_{{\bf k}\sigma}$) is the fermion creation
(destruction) operator for an electron in the state with wave vector
${\bf k}$ and spin projection $\sigma$, and $\xi_{\bf
k}=\varepsilon_{\bf k}-\mu$ with the normal state tight binding
dispersion $\varepsilon_{\bf
k}=-2t[\cos(k_{x})+\cos(k_{y})]-4t_{1}\cos(k_{x})\cos(k_{y})-2t_{2}[\cos(2k_{x})+\cos(2k_{y})]
-4t_{3}[\cos(2k_{x})\cos(k_{y})+\cos(k_{x})\cos(2k_{y})]-4t_{4}\cos(2k_{x})\cos(2k_{y})$
where $(t,t_{1},t_{2},t_{3},t_{4},\mu)=(120,-60,34,7,20,-82)$ with
the unit of meV at 0.11 doping\cite{shan} reproduce the underlying
FS as inferred from recent ARPES experiment\cite{matsui}. As argued
above, the maximum SC gap is achieved near hot
spots\cite{blumberg,krotkov,manske}, which are located much closer
to the zone diagonal, leading to a non-monotonic behavior of the SC
gap. A good fit of $\Delta_{\bf k}$ to the experimental data is
achieved via the inclusion of a higher harmonic, such that
$\Delta_{\bf k}=\sum_{i=1,3}\Delta_{i}[\cos(ik_{x})-\cos(ik_{y})]/2$
with $\Delta_{1}=5.44$ meV and $\Delta_{3}=-2.34$ meV ensures that
the maximum of $|\Delta_{\bf k}|$ along the FS is located at the hot
spots. Corresponding FS and a non-monotonic gap as a function of the
FS angle have been shown in Fig.4 of Ref.20.

By introducing a two-component Nambu spinor operator, $\Psi_{{\bf
k}}=(c_{{\bf k}\uparrow},c_{-{\bf k}\downarrow}^{\dagger})^{\top}$,
the matrix Green's function $G_{0}$ in the superconducting state can
be written by
\begin{eqnarray}
G_{0}({\bf k},i\omega_{n})=\frac{i\omega_{n}\tau_{0}+\xi_{\bf
k}\tau_{2}+\Delta_{\bf k}\tau_{1}}{(i\omega_{n})^{2}-E^{2}_{\bf k}},
\end{eqnarray}
with $E_{\bf k}=(\xi^{2}_{\bf k}+\Delta^{2}_{\bf k})^{1/2}$ the
quasiparticle spectrum and $\tau_{i}$ being the Pauli spin operator.
The corresponding real-space Green's function is
\begin{eqnarray}
G_{0}(i,j;i\omega_{n})&=&\frac{1}{N}\sum_{\bf k}e^{i\bf k\cdot\bf
R_{ij}}G_{0}({\bf k},i\omega_{n}),
\end{eqnarray}
where $\bf R_{ij}=\bf R_{i}-\bf R_{j}$ with $\bf R_{i}$ being
lattice vector. In the presence of a single-site nonmagnetic
impurity of strength $U_{0}$ located at the origin $r_{i}=0$, the
site dependent Green's function in term of the T-matrix
approach\cite{morr} can be obtained as
\begin{eqnarray}
G(i,j;i\omega_{n})&=&G_{0}(i-j;i\omega_{n})\nonumber\\&+&G_{0}(i;i\omega_{n})T(i\omega_{n})G_{0}(j;i\omega_{n}),
\end{eqnarray}
where
\begin{eqnarray}
T(i\omega_{n})=\frac{U_{0}\tau_{3}}{1-U_{0}\tau_{3}G_{0}(0,0;i\omega_{n})}.
\end{eqnarray}
For the d-wave (with or without a higher harmonic) pairing symmetry,
one can find that the local Green's function
$G_{0}(i,i;i\omega_{n})$ is diagonal. As a result, the diagonal
T-matrix reads
\begin{eqnarray}
T_{11,22}(i\omega_{n})=\frac{\pm U_{0}}{1-U_{0}[G_{0}(0,0;\pm
i\omega_{n})]_{11}}
\end{eqnarray}
where the plus (minus) sign denotes $T_{11}$ ($T_{22}$), giving rise
to a particle- ($\omega_{res}<0$) and hole-like ($\omega_{res}>0$)
resonance state. These resonance states generate the sharp peaks in
the LDOS only in the unitary limit
($\mid\omega_{res}\mid/\Delta_{1}\leq1$) where
$1=U_{0}Re[G_{0}(0,0;\pm \omega_{res})]_{11}$.

Finally, the LDOS which can be measured in the STM experiment is
expressed as
\begin{eqnarray}
N(r,\omega)&=&\sum_{\sigma}N_{\sigma}(r,\omega)
\end{eqnarray}
with spin-resolved LDOS
\begin{eqnarray}
N_{\uparrow}(r,\omega)&=&-\frac{1}{\pi}{\rm
    Im}G_{11}(r,r;\omega+i0^{\dagger}), \\
N_{\downarrow}(r,\omega)&=&\frac{1}{\pi}{\rm
Im}G_{22}(r,r;-\omega-i0^{\dagger}).
\end{eqnarray}
The above equations allow a complete solution as long as the
order-parameter relaxation can be ignored.

\begin{figure} \centering
\includegraphics[width=9cm]{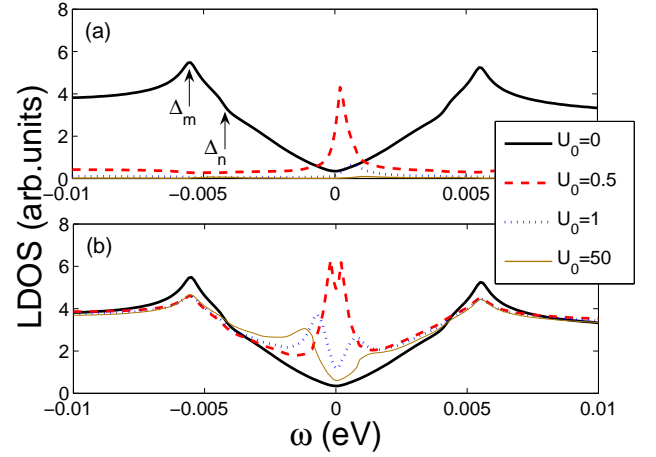}
\vspace{-0.2cm}\caption{\label{Fig1} (Color online) The dependence
of LDOS on scattering strength $U_{0}$ (a) at impurity site and (b)
on the impurity's nearest-neighbor site. $\Delta_{n}$ and
$\Delta_{m}$ denote the antinodal gap and maximum gap at the hot
spot respectively.} \label{fig1}
\end{figure}

In Fig.1 we present the dependence of LDOS on scattering strength
$U_{0}$. The LDOS with $U_{0}=0$ (thick solid line) which is
equivalent to DOS in the clean system, shows two van Hove
singularities at corresponding antinodal gap ($\Delta_{n}$) and
maximum gap at the hot spot ($\Delta_{m}$), indicating the presence
of a non-monotonic $d_{x^{2}-y^{2}}$-wave gap, and is qualitatively
consistent with the recent doping dependence of tunneling spectrum
in Pr$_{2-x}$Ce$_{x}$CuO$_{4-\delta}$\cite{dagan}. In the presence
of a non-magnetic impurity,  a $single$ resonance state appears at
the impurity site (Fig.1(a)). With increasing $U_{0}$, the position
of the resonance state shifts to positive high energy; meanwhile the
spectral weight gradually reduces (the LDOS in Fig.1(a) with
$U_{0}=50eV$ (thin solid line) has been amplified 500 times) and
finally vanishes in the limit $U_{0}\longrightarrow \propto$. In the
LDOS on the impurity's nearest-neighbor site (Fig.1(b)), there are
$two$ resonance states locating at the positive and negative energy
with different spectral weight due to the particle-hole asymmetry.
These features can be understood from the Eq. (3) and Eq. (5), where
the correction to $G(i,i;i\omega_{n})$ due to impurity scattering
reads
\begin{eqnarray}
\delta
G_{11}(i;i\omega_{n})&=&\frac{U_{0}[G_{0}(i;i\omega_{n})]^{2}_{11}}{1-U_{0}[G_{0}(i;i\omega_{n})]_{11}}
\nonumber\\&-&\frac{U_{0}[G_{0}(i;i\omega_{n})]^{2}_{12}}{1-U_{0}[G_{0}(i;-i\omega_{n})]_{11}}.
\end{eqnarray}
At the impurity site, the fact $[G_{0}(i;i\omega_{n})]_{12}=0$ leads
to a single resonance state; while on the impurity's
nearest-neighbor site, spectral weight of both resonance states is
nonzero, $i.e.$ $[G_{0}(i;i\omega_{n})]_{11,12}\neq0$ give rise to
two resonance states. Due to the same spin component at single site,
the spin-resolved LDOS ($N_{\uparrow}$ and $N_{\downarrow}$)
degenerates, resulting in a degenerate $single$ resonance state at
the impurity site and $two$ resonance states on the impurity's
nearest-neighbor site in total LDOS. These features, which are
qualitatively similar to the $d_{x^{2}-y^{2}}$-wave hole-doped
cuprates\cite{morr}, indicate that the inclusion of a higher
harmonic in the gap function basically can not alter the local
electronic structure near a non-magnetic impurity, although it
generates a non-monotonic $d_{x^{2}-y^{2}}$-wave gap in
electron-doped
  cuprate superconductors.


We now compare the above results with those resulted from the
coexisting AF SDW and SC phase. It is convenient to introduce a
$4\times4$ matrix formulation, take four-component Nambu spinor
$\varphi_{{\bf k}}=(c_{{\bf k}\uparrow},c_{{\bf
k+Q}\uparrow},c_{-{\bf k}\downarrow}^{\dagger},c_{-{\bf
k-Q}\downarrow}^{\dagger})^{\top}$ with ${\bf Q}=(\pi,\pi)$ being
the nesting vector, and then write the mean-field Hamiltonian as $
H_{SC+SDW}=\sum_{{\bf k}}\varphi^{+}_{{\bf k}}(\xi_{\bf
k}\tau_{3}\rho_{0}+M\tau_{1}\rho_{0}+\Delta_{\bf
k}\tau_{3}\rho_{1})\varphi_{{\bf k}}$, where $\tau_{3}\rho_{1}=
\left (\matrix{0 &\tau_{3}\cr \tau_{3} &0\cr}\right)$, $M$ is AF SDW
order parameter and $\Delta_{\bf
k}=\Delta_{1}[\cos(k_{x})-\cos(k_{y})]/2$ is monotonic
$d_{x^{2}-y^{2}}$-wave SC gap. Note that from now on the wave vector
{\bf k} is restricted to the magnetic Brillouin zone (MBZ).

The relevant matrix Green's function is obtained as
\begin{eqnarray}
g^{-1}_{0}({\bf k},i\omega_{n})=i\omega_{n}-\xi_{\bf
k}\tau_{3}\rho_{0}-M\tau_{1}\rho_{0}-\Delta_{\bf k}\tau_{3}\rho_{1}.
\end{eqnarray}
To solve for the resonance state in the coexisting AF SDW and SC
phase, we define the $2\times2$ Green's function as
\begin{eqnarray}
G_{0}(i,j;i\omega_{n})&=&\frac{1}{N}\sum_{\bf k}e^{i\bf k\cdot\bf
R_{ij}}\nonumber\\
&\times& \left (\matrix{G^{1}_{0}({\bf k},i\omega_{n})
&G^{2}_{0}({\bf k},i\omega_{n})\cr G^{3}_{0}({\bf k},i\omega_{n})
&G^{4}_{0}({\bf k},i\omega_{n})\cr}\right),
\end{eqnarray}
where
\begin{eqnarray}
G^{{\bf I}}_{0}({\bf k},i\omega_{n})&=&e^{-i\bf Q\cdot\bf
R_{j}}[g_{0}]_{{\bf I}}^{12}({\bf k},i\omega_{n})+e^{i\bf Q\cdot\bf
R_{i}}[g_{0}]_{{\bf I}}^{21}({\bf
k},i\omega_{n})\nonumber\\&+&e^{i\bf Q\cdot\bf R_{ij}}[g_{0}]_{{\bf
I}}^{22}({\bf k},i\omega_{n})+[g_{0}]_{{\bf I}}^{11}({\bf
k},i\omega_{n}),
\end{eqnarray}
with ${\bf I}=1,2,3,4$ denoting the left-top, right-top, left-bottom
and right-bottom $2\times2$ block element of $g_{0}({\bf
k},i\omega_{n})$. Applying T-matrix approach\cite{morr}, we can
easily get the LDOS in the presence of a non-magnetic impurity.

The dependence of LDOS on scattering strength $U_{0}$ for different
AF SDW order $M$ is plotted in Fig.2. Following the discussions in
Ref.19, we take the independent particle dispersion $\xi_{\bf k}$,
and consider the doping dependent AF SDW order $M$. We in the
following calculation choose the self-consistent value $M=0.14eV$ at
0.11 doping\cite{shan}, and the decreasing $M$ values corresponding
to the doping increasing .

Before considering the effect of the impurity we briefly review the
evolution of DOS with AF SDW order in the SC state\cite{liu}. In
Fig.2 the LDOS with $U_{0}=0$ (thick solid line) is equivalent to
DOS in the clean system. In the limit $M=0eV$, as seen in hole-doped
cuprates\cite{morr}, the DOS (thick solid line in Fig.2(a)) at low
energies behaves to be V-shaped like with a monotonic
$d_{x^{2}-y^{2}}$-wave SC gap, and a coherent peak locates at the
maximal gap edge. After introducing AF SDW order, another coherent
peak appears at the energy less than the maximum gap (thick solid
line in Fig.2(b)). Thus, a non-monotonic $d_{x^{2}-y^{2}}$-wave SC
gap occurs in coexisting AF SDW and SC state. In particular, the DOS
shows the U-shaped behavior at sufficiently large SDW order
$M=0.14eV$ with doping $x=0.11$\cite{shan} (thick solid line in
Fig.2(c)), which has been observed in earlier point contact
tunneling spectra\cite{alff,biswas}. These unusual evolutions of DOS
with AF SDW order are qualitatively similar to the doping evolution
of DOS\cite{yuan}, and have been explained as the result of the
coexisting AF SDW and SC state\cite{yuan,liu}.

\begin{figure} \centering
\includegraphics[width=9cm]{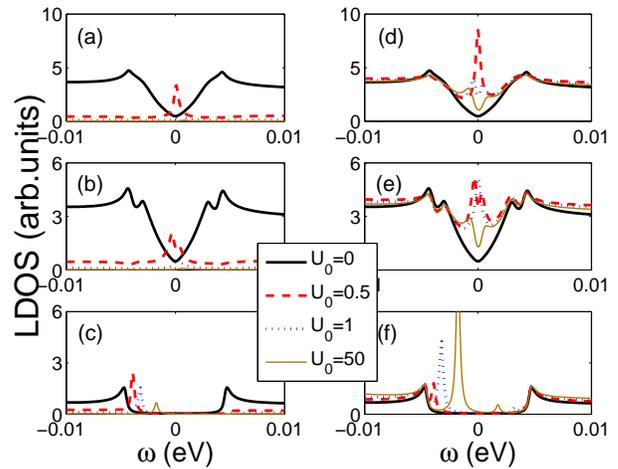}
\vspace{-0.2cm}\caption{\label{Fig2} (Color online) The dependence
of LDOS on scattering strength $U_{0}$ for different AF SDW order
$M$. LDOS at impurity site: (a)$M=0eV$, (b)$M=0.05eV$, and
(c)$M=0.14eV$; and  LDOS on the impurity's nearest-neighbor site:
(d)$M=0eV$, (e)$M=0.05eV$, and (f)$M=0.14eV$.} \label{fig2}
\end{figure}
\begin{figure} \centering
\includegraphics[width=9cm]{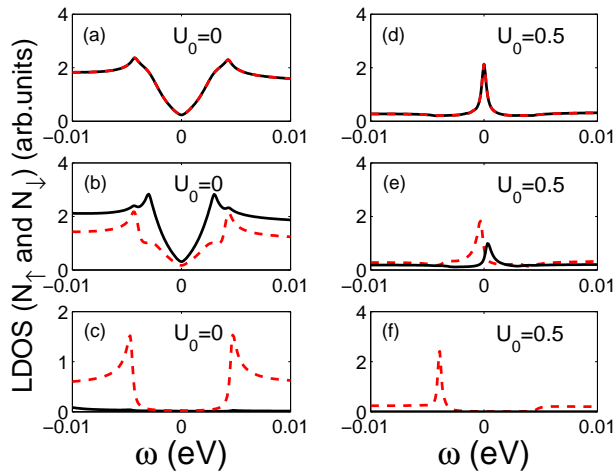}
\vspace{-0.2cm}\caption{\label{Fig3} (Color online) The
spin-resolved LDOS $N_{\uparrow}$ (dashed line) and $N_{\downarrow}$
(solid line) at impurity site without scattering $U_0=0eV$ for AF
SDW gap (a)$M=0eV$, (b)$M=0.05eV$,and (c)$M=0.14eV$; and with
$U_0=0.5eV$ for (d)$M=0eV$, (e)$M=0.05eV$,and (f)$M=0.14eV$.}
\label{fig3}
\end{figure}
%
%

We proceed to analyze the dependence of LDOS on scattering strength
$U_{0}$ near a non-magnetic impurity. For the limit AF SDW order
$M=0eV$, we show that the LDOS at the impurity site (Fig.2(a)) and
on the impurity's nearest-neighbor site (Fig.2(d)) are similar to
that obtained from the $d_{x^{2}-y^{2}}$-wave hole-doped
superconductors \cite{morr}. With increasing AF SDW order, the LDOS
at the impurity site are qualitatively different. In Fig.2(b), it is
clearly shown that two resonance states at the impurity site occur
near the Fermi energy, which indicates that the degenerate $single$
resonance state with $M=0$ has separated due to the presence of AF
SDW order. In principle, the spin-resolved LDOS should give rise to
multiple resonance states on the impurity's nearest-neighbor site,
though they are actually not easy to be resolved in Fig.2(e) because
of the resonance states crossing each other near Fermi energy. At
sufficiently large SDW order $M=0.14eV$, one resonance state at the
impurity site (Fig.2(c)) exists and shifts towards the gap edge, the
other one is barely visible due to the vanishing spectral weight.
For the better understanding of such important features, the
spin-resolved LDOS ($N_{\uparrow}$ and $N_{\downarrow}$) at impurity
site without scattering $U_0=0eV$ and with scattering strength
$U_0=0.5eV$ are shown in Fig.3 for increasing $M$. When $M=0$, the
spin-resolved $N_{\uparrow}$ and $N_{\downarrow}$ with $U_0=0eV$
degenerate (Fig.3a), the resulting resonance state in spin-resolved
LDOS $N_{\uparrow}$ and $N_{\downarrow}$ with $U_0=0.5eV$ are
located at the same resonance energy, leading to a degenerate
$single$ resonance state (Fig.3d). With $M$ increasing, the LDOS
$N_{\uparrow}$ is not equal to $N_{\downarrow}$ (Fig.3b), thus the
degenerate spin-resolved LDOS separates, leading to two resonance
states at impurity site with different spectral weight (Fig.3e). At
sufficiently large SDW order M, the LDOS $N_{\uparrow}$ is dominant
over $N_{\downarrow}$ (Fig.3c), thus one single resonance state from
spin-up component exists and shifts towards the gap edge due to the
U-shaped DOS (Fig.3f), and the other one from spin-down component is
barely visible due to the vanishing spectral weight (solid line in
Fig.3c). Therefore due to the different spin components induced by
the presence of AF SDW gap, an existing impurity will be polarized
by a local net spin-up or spin-down, which leads to the splitting of
the LDOS. These unique features do not appear in the scenario of
$d_{x^{2}-y^{2}}$-wave superconductivity with a higher harmonic and
should be detected by the spin-polarized STM measurement.


In summary, we analyze the LDOS around a non-magnetic impurity in
electron-doped cuprate superconductors starting from two different
scenarios: $d_{x^{2}-y^{2}}$-wave superconductivity with a higher
harmonic versus $d_{x^{2}-y^{2}}$-wave superconductivity coexisting
with AF SDW order. We find that in both cases the DOS indicates the
presence of non-monotonic $d_{x^{2}-y^{2}}$-wave gap, qualitatively
consistent with the recent tunneling spectrum measurement in
Pr$_{2-x}$Ce$_{x}$CuO$_{4-\delta}$\cite{dagan}, therefore both of
them have been thought to be the possible physical origins of the
non-monotonic $d_{x^{2}-y^{2}}$-wave gap. We also find that the
inclusion of a higher harmonic basically doesn't alter the local
electronic structure near a non-magnetic impurity; in contrast, with
introducing AF SDW order, the LDOS presents spin-resolved feature,
$i.e.$ a degenerate $single$ resonance state at the impurity site
separates into two resonance states due to the different spin
component induced by the presence of AF SDW order. Thus we strongly
suggest that the future spin-polarized STM measurements should be
performed to differentiate two above scenarios and shed light on the
real physical origin of non-monotonic $d_{x^{2}-y^{2}}$-wave gap in
electron-doped cuprate superconductors.

We thank Prof. I. Eremin, Prof. Xi Dai for the careful reading of
the manuscript and fruitful discussions.



\end{document}